\documentclass[twocolumn,amsmath,amssymb,prm,aps,showkeys]{revtex4}
\usepackage{caption}
\usepackage{ragged2e}
\usepackage{graphicx}
\usepackage{dcolumn}
\usepackage{bm}
\usepackage{xcolor}
\usepackage{hyperref}
\usepackage{booktabs}
\usepackage{multirow}
\usepackage{array}
\usepackage{physics}
\usepackage{siunitx}
\usepackage{float}
\usepackage{mathtools}
\usepackage{braket}
\usepackage{enumitem}
\makeatletter
\long\def\@makecaption#1#2{%
	\par
	\vskip\abovecaptionskip
	\begingroup
	\small\rmfamily
	\justifying
	\noindent
	#1\quad #2\par
	\endgroup
	\vskip\belowcaptionskip
}
\makeatother
\begin{document}


\title{Coupled Altermagnetic and Valley Physics in Monolayer VBiSe$_3$}


\author{Divyanshi Tyagi}
\email{divyanshi.tyagi@physics.iitd.ac.in}
\affiliation{Department of Physics, Indian Institute of Technology Delhi, New Delhi 110016, India}

\author{Saswata Bhattacharya}
\email{saswata@physics.iitd.ac.in}
\affiliation{Department of Physics, Indian Institute of Technology Delhi, New Delhi 110016, India}


\begin{abstract}
	
The interplay of compensated magnetism, spin--orbit coupling (SOC), and
valley degrees of freedom provides a promising route toward spin--valley
control without conventional ferromagnetism. Here, using first-principles
calculations combined with Wannier-based analysis, we reveal coupled
altermagnetic and valley physics in monolayer VBiSe$_3$. In the absence of
SOC, the compensated N\'eel antiferromagnetic state, whose opposite-spin V
sublattices are related only by mirror operations, exhibits
momentum-dependent spin splitting with a characteristic $i$-wave
altermagnetic texture. Upon inclusion of SOC, the conduction-band states
develop anisotropic Rashba-like spin splitting together with a sizable
$K/K'$ valley splitting of about $112.2$~meV. Orbital analysis and an
effective SOC Hamiltonian identify the hybridization between
SOC-active Se-$p_{x,y}$ states and predominantly V-$d_{z^2}$
conduction-band states as an important microscopic origin of the valley
splitting. The two valleys carry Berry curvature of the same sign but
different magnitude, unlike the opposite-sign pattern of conventional
valley Hall systems, so the large valley splitting, rather than a valley
Hall effect, offers a possible route to valley-selective carrier
occupation by gating. Moreover, a modest $-1\%$ biaxial compressive strain
drives the system into a quasi-half-valley-metallic state with
valley-polarized electrons and non-valley-polarized holes. These results
establish monolayer VBiSe$_3$ as a promising platform for coupled
altermagnetic, spin--orbit, and valley phenomena with strain-tunable
electronic states.
\end{abstract}


\maketitle

\section{Introduction}

Valleys, corresponding to local extrema of the conduction or valence
bands at inequivalent points in momentum space, provide an additional
degree of freedom for information encoding beyond charge and spin
~\cite{Schaibley2016,Xiao2012,Vitale2018,Mak2014}. In two-dimensional
(2D) materials with hexagonal Brillouin zones, the inequivalent $K$ and
$K'$ valleys can possess distinct spin, orbital, and geometrical
properties. A central quantity governing such valley physics is the Berry
curvature, which acts as an effective magnetic field in momentum space
and produces anomalous transverse carrier motion under an applied
electric field
~\cite{Xiao2007Valley,Xiao2010,Nagaosa2010AHE,Olsen2015}. In
noncentrosymmetric systems, spin--orbit coupling (SOC) further couples the
spin, orbital, and valley degrees of freedom, enabling valley-dependent
spin textures and transverse responses~\cite{Xiao2012,Schaibley2016}.
Lifting the energetic degeneracy between opposite valleys is therefore
a key step toward exploiting the valley degree of freedom, since it allows
one valley to be populated preferentially, motivating extensive searches
for magnetic and ferrovalley materials with spontaneous valley polarization
~\cite{Tong2016,Li2020PRM,Peng2021PRB,LiAn2022PRM}.

Magnetic 2D materials provide a natural route toward such spontaneous
valley polarization because broken time-reversal symmetry can remove the
energetic equivalence between opposite valleys without an external
magnetic field~\cite{Xu2021PRB,Peng2021PRB,Ahammed2022PRB,
	Peng2020Nb3I8,Cui2021VSi2N4}. Antiferromagnets are particularly attractive
because their compensated magnetic moments suppress stray fields while
retaining the possibility of ultrafast spin dynamics
~\cite{Jungwirth2016,Baltz2018}. Recent predictions of spontaneous valley
polarization and anomalous valley responses in 2D antiferromagnets further
demonstrate the potential of compensated magnetic systems for
valleytronics~\cite{Guo2024Fe2CF2}. In this context, altermagnetism has
introduced a distinct form of compensated magnetism in which a vanishing
net magnetic moment coexists with pronounced nonrelativistic
momentum-dependent spin splitting
~\cite{Smejkal2022Phase,Smejkal2022Landscape}. Depending on the magnetic
crystal symmetry, the resulting spin-splitting patterns can exhibit
characteristic $d$-, $g$-, or $i$-wave forms
~\cite{Zeng2024PRB,Roig2024PRB}. This raises an important question of how
such nonrelativistic altermagnetic spin splitting evolves when coupled to
SOC and valley degrees of freedom in two dimensions. Recent studies have
shown that breaking $\mathcal{PT}$ symmetry in N\'eel antiferromagnets can
generate $i$-wave altermagnetic spin splitting together with SOC-driven
valley polarization~\cite{Sun2025VPSe3}, highlighting the possibility of coupling altermagnetic spin splitting
with valley polarization in two dimensions.

The coexistence of altermagnetic order and SOC in a noncentrosymmetric
crystal provides an additional route for manipulating the band-edge
states. SOC can generate Rashba-type spin--momentum locking
~\cite{Bihlmayer2022,Miyamoto2018PRB,Manchon2015Rashba}, reconstruct the
spin texture, and strongly modify the Berry-curvature distribution across
the Brillouin zone. The magnetic-moment orientation can further influence
the magnitude and even the occurrence of spontaneous valley polarization
in 2D magnetic systems~\cite{Gao2025VInSe3}. Despite these developments,
the simultaneous occurrence of nonrelativistic altermagnetic spin
splitting, SOC-induced Rashba-like behavior, spontaneous valley
polarization, and an inequivalent valley-resolved Berry-curvature distribution remains
comparatively unexplored. Moreover, because the valley states are highly
sensitive to lattice deformation, strain offers an effective means of
reorganizing the band-edge hierarchy and driving transitions toward
valley-selective metallic and half-valley-metallic phases
~\cite{Zang2021,Li2023,Sheng2022RuCl2,Huan2021MBr2}.

In this work, we investigate monolayer VBiSe$_3$, a noncentrosymmetric
2D antiferromagnet with $P31m$ crystal symmetry and $C_{3v}$ point-group
symmetry. In the absence of SOC, the opposite-spin V sublattices of the
N\'eel antiferromagnetic (AFM) state are related only by mirror operations,
which allows an $i$-wave momentum-dependent spin-splitting texture.
Including SOC substantially reconstructs the conduction-band states,
producing anisotropic Rashba-like spin splitting together with a sizable
$K/K'$ valley splitting. The two valleys carry Berry curvature of the same
sign but different magnitude, so, unlike in conventional valley Hall
systems, they are not deflected in opposite directions. The valley-related
functionality instead rests on the large energy splitting, which permits
valley-selective occupation. Furthermore, a modest -$1\%$ biaxial compressive
strain drives the system into a quasi-half-valley-metallic (QHVM) state. 

\section{Computational Details}

First-principles calculations were performed within density functional
theory (DFT) using the projector augmented-wave (PAW) method implemented
in the Vienna \textit{Ab initio} Simulation Package (VASP)
~\cite{Kresse1996a,Kresse1996b,Blochl1994,Kresse1999}. The
exchange--correlation interaction was treated within the generalized
gradient approximation using the Perdew--Burke--Ernzerhof (PBE)
functional~\cite{Perdew1996}. A plane-wave kinetic-energy cutoff of
550~eV and a $\Gamma$-centered $9\times9\times1$  $k$-point mesh were used
for Brillouin-zone sampling. A vacuum layer of 25~\AA{} was introduced
along the out-of-plane direction to suppress interactions between
periodic images. All structures were fully relaxed until the
Hellmann--Feynman forces on each atom were smaller than
$10^{-6}$~eV\,\AA$^{-1}$.

The localized V-$3d$ electrons were treated within the rotationally
invariant PBE+$U$ approach of Dudarev \textit{et al.}
~\cite{Dudarev1998}, using an effective Hubbard parameter
$U_{\mathrm{eff}}=3.2$~eV \cite{Sun2025VPSe3}. SOC was included self-consistently in the relativistic electronic
structure, valley-splitting, spin-texture, and magnetic-anisotropy
calculations. The magnetic anisotropy energy (MAE) was obtained from the
difference between the SOC total energies for in-plane and out-of-plane
magnetic orientations. Dynamical stability was examined from the phonon
dispersion calculated using \textsc{PHONOPY} with a $3\times3\times1$
supercell~\cite{Togo2023}.

The nonrelativistic momentum-dependent spin splitting was obtained from
the spin-resolved electronic structure calculated without SOC and mapped
in momentum space to characterize the $i$-wave spin-splitting texture.
The SOC-induced conduction-band spin texture was evaluated from the
momentum-resolved spin expectation values. Maximally localized Wannier
functions were constructed
using \textsc{Wannier90}~\cite{Mostofi2008,Mostofi2014}. 

\section{Results and Discussion}

\subsection*{Crystal symmetry and magnetic ground state}

\begin{figure}[htp]
	\centering
	\includegraphics[width=0.5\textwidth]{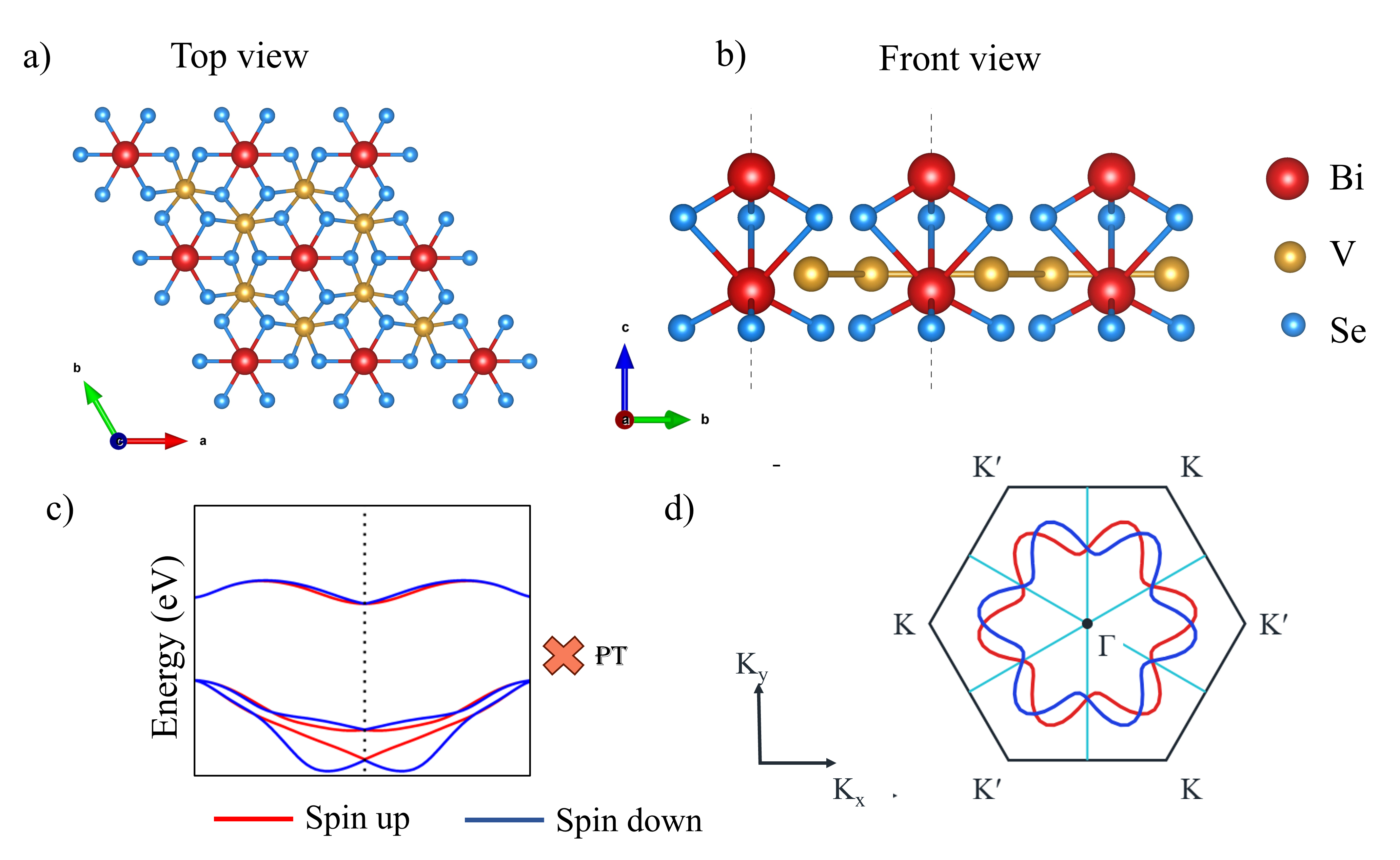}
	\caption{
		(a) Top and (b) front views of monolayer VBiSe$_3$.
		(c) Schematic spin-resolved electronic bands in the absence of
		spin--orbit coupling (SOC), illustrating the momentum-dependent
		separation between the opposite spin channels in the absence of
		$\mathcal{PT}$ symmetry.
		(d) Hexagonal Brillouin zone showing the relevant high-symmetry points.
	}
	\label{fig1}
\end{figure}

Figures~\ref{fig1}(a) and \ref{fig1}(b) show the top and front views of monolayer VBiSe$3$, respectively. The primitive unit cell consists of two V, two Bi, and six Se atoms and adopts a hexagonal noncentrosymmetric structure with space group $P31m$ (No.~157) and point group $C{3v}$. The optimized in-plane lattice constants are $a=b=6.36$~\AA. The corresponding hexagonal Brillouin zone and relevant high-symmetry points are shown in Fig.~\ref{fig1}(d).

Four representative magnetic configurations, namely ferromagnetic (FM), N'eel antiferromagnetic (AFM-N'eel), AFM-stripy, and AFM-zigzag, were considered, as shown in Sec.~I of the Supplemental Material (SM). Among these configurations, the AFM-N'eel state is found to be the magnetic ground state. Its energy is lower than those of the FM, AFM-stripy, and AFM-zigzag configurations by 208, 134.18, and 646~meV per unit cell, respectively, establishing the energetic preference for N'eel-type AFM ordering in monolayer VBiSe$_3$.

The two V atoms carry opposite magnetic moments of approximately $+1.964$ and $-1.964~\mu_{\mathrm B}$, respectively, resulting in a vanishing net magnetization. To establish the altermagnetic character of this magnetic configuration, we examined the symmetry operations of the crystal structure within the nonrelativistic spin-group formalism. The threefold rotational operations preserve the individual V sublattices, whereas the three vertical mirror operations interchange the oppositely magnetized V sites. When combined with a twofold spin rotation $C_2$ perpendicular to the collinear spin axis, these mirror operations give rise to the spin-group symmetries $[C_2\Vert M_i]$. Importantly, the opposite-spin sublattices are not connected by inversion or pure lattice translation. Furthermore, the degeneracy-enforcing spin-group operations associated with $M_z$ and $C_{2z}$ are absent. Consequently, the magnetic symmetry permits momentum-dependent nonrelativistic spin splitting, providing the symmetry basis for altermagnetic behavior in monolayer VBiSe$_3$.

The preferred orientation of the magnetic moments was examined through the magnetocrystalline anisotropy energy (MAE), defined as

\begin{equation}
\mathrm{MAE}=E_{x/y}-E_z,
\end{equation}

where $E_{x/y}$ and $E_z$ denote the total energies for in-plane and out-of-plane spin orientations, respectively. The calculated MAE of 1.18~meV per unit cell identifies the out-of-plane direction as the magnetic easy axis. This magnetic configuration was considered in the subsequent SOC calculations. The dynamical stability of the monolayer was further examined through its phonon dispersion, which exhibits no significant imaginary frequencies throughout the Brillouin zone (Fig.~S2).

\subsection*{Electronic structure and spin-valley coupling}

\begin{figure}[t]
	\centering
	\includegraphics[width=0.5\textwidth]{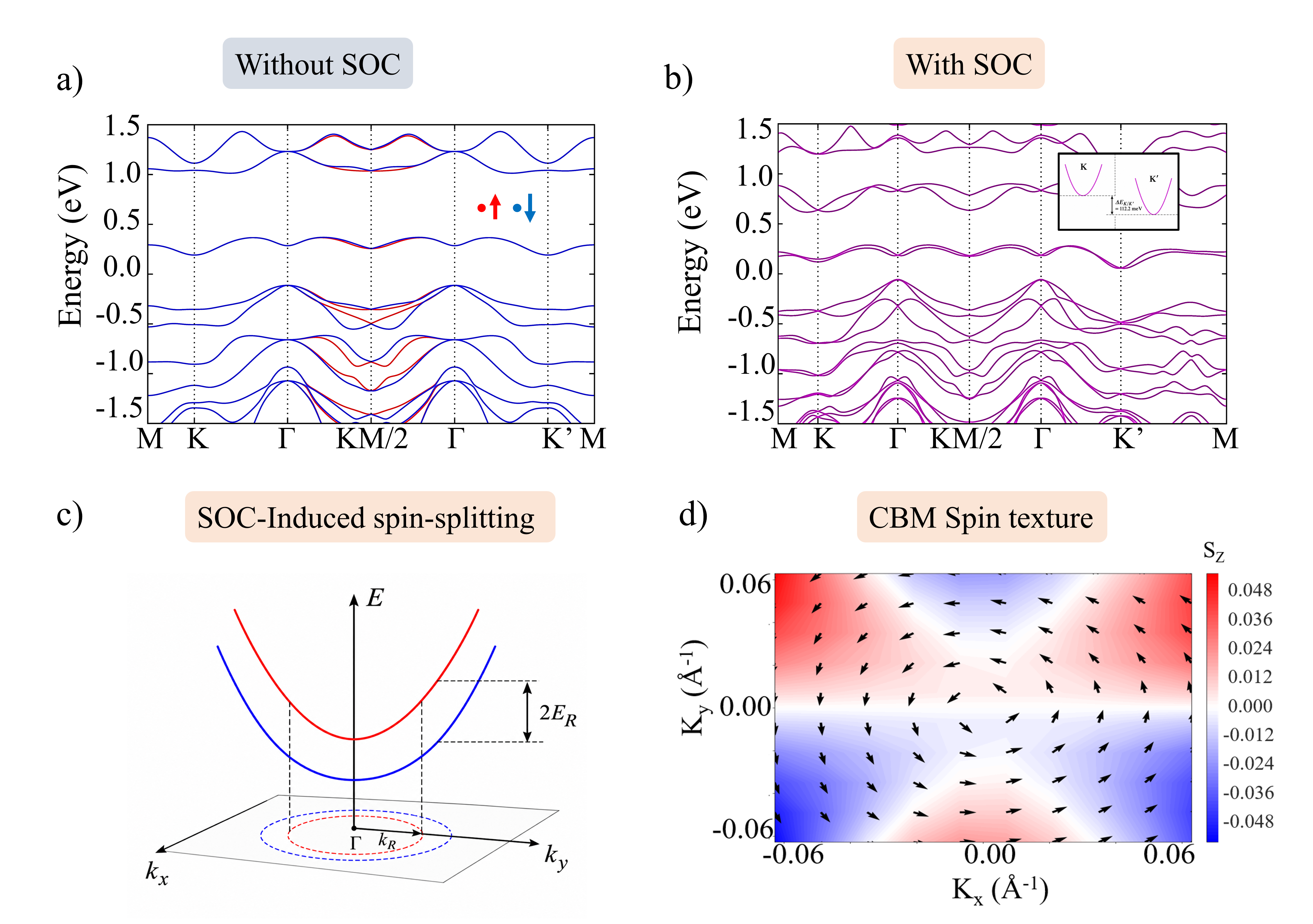}
	\caption{
		Electronic structure of monolayer VBiSe$_3$:
		(a) spin-resolved band structure without SOC and
		(b) band structure with SOC. The inset highlights the
		energetic splitting between the conduction-band valleys at $K$ and $K'$.
		(c) Schematic representation of the SOC-induced Rashba-like spin
		splitting, indicating the momentum offset $k_R$ and Rashba energy $E_R$.
		(d) SOC-induced conduction-band spin texture in the $k_x$--$k_y$ plane
		at $k_z=0$. Arrows denote the in-plane spin components, while the
		color scale represents $S_z$.
	}
	\label{fig2}
\end{figure}

Having established the compensated AFM ground state and its underlying magnetic symmetry, we next investigate the electronic structure of monolayer VBiSe$_3$. Figure~\ref{fig2}(a) shows the spin-resolved band structure calculated without SOC, revealing finite spin splitting despite the vanishing net magnetization. The persistence of this splitting in the absence of SOC indicates its nonrelativistic, exchange-driven origin, consistent with the altermagnetic character of the system~\cite{Smejkal2022Phase,Smejkal2022Landscape}.

Upon inclusion of SOC, the electronic structure undergoes a pronounced reconstruction, as shown in Fig.~\ref{fig2}(b). Monolayer VBiSe$_3$ is an indirect-gap semiconductor with a band gap of 0.304~eV. The conduction-band minima occur at the inequivalent $K$ and $K'$ valleys, whereas the valence-band maximum is located at the $\Gamma$ point. The inclusion of SOC lifts the energetic equivalence of the two conduction-band valleys. This valley splitting is defined as

\begin{equation}
\Delta E_{\mathrm{val}}^{\mathrm{CBM}}
=
\left|E_{\mathrm{C}}(K)-E_{\mathrm{C}}(K')\right|,
\end{equation}

where $E_{\mathrm{C}}(K)$ and $E_{\mathrm{C}}(K')$ denote the conduction-band energies at the two valleys. For the pristine monolayer, we obtain

\begin{equation}
\Delta E_{\mathrm{val}}^{\mathrm{CBM}}
=
112.2~\mathrm{meV}.
\end{equation}

This sizable valley splitting indicates pronounced energetic inequivalence between the two conduction-band valleys. Notably, the calculated value exceeds the spontaneous valley splittings reported for several representative two-dimensional magnetic valley systems, including VPSe$_3$ ($2.91$~meV), VInSe$_3$ ($23.4$~meV), LaBr$_2$ ($33$~meV), TiVI$_6$ ($22$~meV), and VAgP$_2$Se$_6$ ($15$~meV)~\cite{Sun2025VPSe3,Gao2025VInSe3,Zhao2019LaBr2,Du2020TiVI6,Song2018VAgP2Se6}. The valley inequivalence obtained in the SOC-included electronic structure reflects the interplay of magnetic order, broken inversion symmetry, and relativistic SOC.

The SOC-induced modification of the conduction bands is also accompanied by pronounced momentum-dependent spin splitting, as schematically illustrated in Fig.~\ref{fig2}(c). The strength of the Rashba-like splitting can be characterized by the effective parameter

\begin{equation}
\alpha_R=\frac{2E_R}{k_R},
\end{equation}

where $E_R$ and $k_R$ denote the Rashba energy and momentum offset, respectively. The calculated value is $\alpha_R\approx0.79~\mathrm{eV\,\AA}$. The corresponding conduction-band spin texture is shown in Fig.~\ref{fig2}(d), where the arrows represent the in-plane spin components and the color scale denotes the out-of-plane component $S_z$. The spin texture exhibits a pronounced anisotropy and deviates from the ideal isotropic Rashba configuration, reflecting the combined influence of SOC, antiferromagnetic exchange, and the noncentrosymmetric $C_{3v}$ crystal environment.

\begin{figure*}[htp]
	\centering
	\includegraphics[width=1\textwidth]{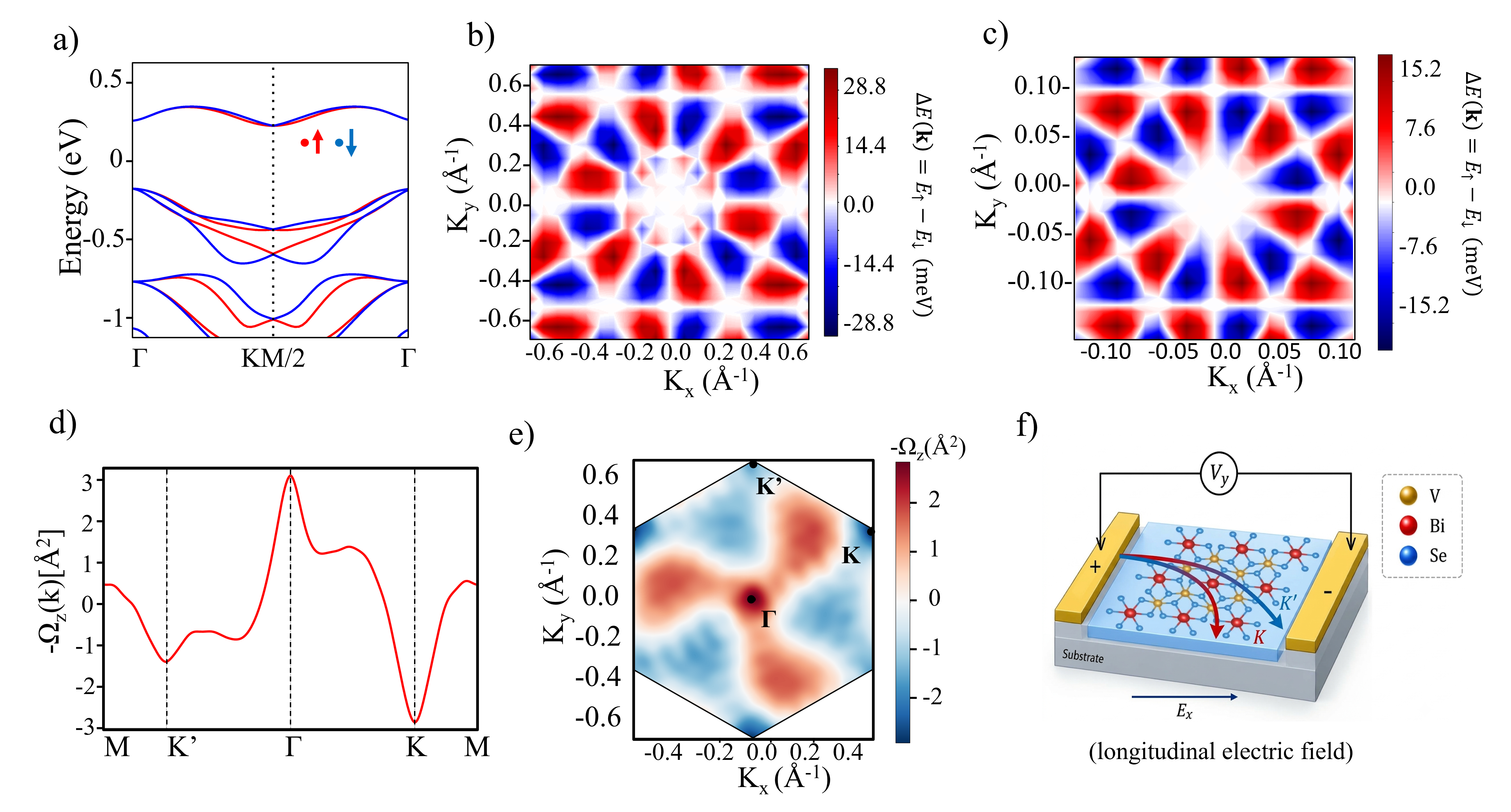}
	\caption{
		Momentum-space spin splitting and Berry-curvature response of monolayer
		VBiSe$_3$.
		(a) Enlarged spin-resolved band structure calculated without SOC,
		illustrating the nonrelativistic splitting between the opposite spin
		channels.
		(b) Momentum-space distribution of the spin splitting
		$\Delta E(\mathbf{k})=E_{\uparrow}(\mathbf{k})-E_{\downarrow}(\mathbf{k})$
		over the extended $k$-space region.
		(c) Enlarged view of the spin-splitting distribution around the central
		momentum region, revealing the characteristic $i$-wave pattern.
		(d) Berry curvature $\Omega_z(\mathbf{k})$ calculated along the
		high-symmetry path.
		(e) Two-dimensional momentum-space distribution of the Berry curvature.
        (f) Schematic device geometry illustrating the unidirectional, velocity-imbalanced anomalous transverse deflection of $K$- and $K'$-valley carriers ($v_K > v_{K'}$) under an applied longitudinal electric field $E_x$.
	}
	\label{fig3}
\end{figure*}

To further characterize the nonrelativistic spin splitting, Fig.~\ref{fig3}(a) presents an enlarged view of the spin-resolved bands calculated without SOC. The momentum-dependent energy separation between the opposite-spin channels is quantified by

\begin{equation}
\Delta E(\mathbf{k})=
E_{\uparrow}(\mathbf{k})-E_{\downarrow}(\mathbf{k}).
\end{equation}

The corresponding two-dimensional distribution of $\Delta E(\mathbf{k})$ is shown in Fig.~\ref{fig3}(b), while Fig.~\ref{fig3}(c) provides an enlarged view of the central momentum region. The spin splitting exhibits alternating signs across different momentum-space sectors, giving rise to a characteristic sixfold pattern consistent with an $i$-wave-like spin-splitting texture. Together with the compensated AFM ground state and the magnetic symmetry discussed above, these results demonstrate the nonrelativistic, momentum-dependent spin splitting characteristic of altermagnetic ordering in monolayer VBiSe$_3$.

\subsection*{SOC Hamiltonian}

The orbital-projected electronic structure (see SM sec III) shows
that the valence-band maximum is mainly composed of Se-$p_x$ and Se-$p_y$
states with additional V-$d$ contributions, whereas the conduction-band
minimum (CBM), where the pronounced $K/K'$ valley splitting occurs, is
dominated by the V-$d_{z^2}$ orbital with a finite admixture of
Se-$p_{x,y}$ character. The latter contribution is particularly important
because complex combinations of the $p_x$ and $p_y$ orbitals carry finite
orbital angular momentum and can therefore couple efficiently to SOC.

Similar SOC-based descriptions have been employed to analyze
valley-dependent energy shifts in related 2D magnetic
systems~\cite{Sun2025VPSe3,Gao2025VInSe3}. The atomic SOC interaction
is described by~\cite{Dai2008SOC,Whangbo2015SOC}

\begin{equation}
\hat{H}_{\mathrm{SOC}}
=
\lambda\hat{\mathbf{S}}\cdot\hat{\mathbf{L}},
\end{equation}

where $\lambda$ is the SOC strength, while $\hat{\mathbf{S}}$ and
$\hat{\mathbf{L}}$ denote the spin and orbital angular-momentum operators,
respectively. For the out-of-plane magnetic orientation established from
the MAE calculation, and considering the leading spin-conserving
contribution within a fixed spin channel, the SOC Hamiltonian reduces to

\begin{equation}
\hat{H}_{\mathrm{SOC}}^{0}
=
\lambda\hat{S}_{z}\hat{L}_{z}.
\end{equation}

The general angular dependence of the SOC Hamiltonian and the separation
of the spin-conserving and spin-flip contributions are provided in
Sec.~IV of the SM.

To illustrate the SOC-active contribution from the Se-$p_{x,y}$ states,
the corresponding component of the conduction-band valley states can be
represented by

\begin{equation}
|\phi_{c}^{\tau}\rangle
=
\frac{1}{\sqrt{2}}
\left(
|p_x\rangle+i\tau|p_y\rangle
\right),
\end{equation}

where $\tau=+1$ and $-1$ denote the $K$ and $K'$ valleys, respectively.
These complex orbital combinations carry opposite orbital-angular-momentum
characters and therefore acquire valley-dependent SOC energy corrections.
Defining
$\beta=\langle\uparrow|\hat{S}_{z}|\uparrow\rangle$, the corresponding
splitting of the idealized SOC-active $p$-orbital contribution is

\begin{equation}
\Delta E_{c}^{(p)}
=
E_{c}^{K}-E_{c}^{K'}
=
2\lambda\beta .
\end{equation}

A detailed derivation of this expression is provided in the SM.

Importantly, the actual CBM of monolayer VBiSe$_3$ is predominantly of
V-$d_{z^2}$ character. For an isolated $d_{z^2}$ orbital, the magnetic
quantum number is $m=0$, giving

\begin{equation}
\langle d_{z^2}|\hat{L}_{z}|d_{z^2}\rangle=0,
\end{equation}

so that a pure $d_{z^2}$ state does not acquire a first-order
valley-dependent shift through the $\hat{L}_z$ term alone. The finite
Se-$p_{x,y}$ admixture therefore introduces SOC-active orbital character
into the predominantly V-$d_{z^2}$ conduction-band states, allowing
different SOC energy corrections at the $K$ and $K'$ valleys. A similar
situation occurs in transition-metal dichalcogenide-based valley systems,
where the conduction-band edge is predominantly of $d_{z^2}$ character,
while smaller chalcogen-$p_{x,y}$ contributions provide SOC-active orbital
character and contribute to the conduction-band spin or valley splitting
~\cite{Tong2016,Yu2015Valley,Zhang2019WSe2CrI3}. The calculated
conduction-band valley splitting should consequently be viewed as a
property of the full hybridized V-$d_{z^2}$/Se-$p_{x,y}$ states rather
than of the individual orbital components alone. This orbital picture
provides a microscopic basis for the strong SOC-induced valley
inequivalence obtained from the first-principles calculations.

\subsection*{Berry curvature}

Berry curvature describes the geometrical properties of Bloch electronic states in momentum space and plays an important role in the anomalous transport properties of solids~\cite{Xiao2007Valley,Xiao2010Berry,Nagaosa2010AHE}. Its distribution is constrained by the underlying crystal and magnetic symmetries. In particular, time-reversal symmetry relates the Berry curvature at opposite momenta with opposite signs, whereas spatial inversion symmetry imposes an equal-sign relation. Since both symmetries are individually broken in monolayer VBiSe$_3$, these conventional symmetry constraints do not establish a definite relationship between the Berry curvatures at the $K$ and $K'$ valleys. Nevertheless, additional constraints arising from the magnetic point group must be considered when interpreting the valley-dependent Berry-curvature distribution.

The Berry curvature was calculated using the Wannier-interpolated Hamiltonian. Within the Kubo formalism, the Berry curvature of the $n$th band is expressed as~\cite{Xiao2010Berry}

\begin{equation}
\Omega_{n,z}(\mathbf{k})
=
-2\hbar^2\,\mathrm{Im}
\sum_{m\neq n}
\frac{
	\langle u_{n\mathbf{k}}|\hat{v}_{x}|u_{m\mathbf{k}}\rangle
	\langle u_{m\mathbf{k}}|\hat{v}_{y}|u_{n\mathbf{k}}\rangle
}{
	\left(E_{m\mathbf{k}}-E_{n\mathbf{k}}\right)^2
}.
\end{equation}

Here, $E_{n\mathbf{k}}$ and $|u_{n\mathbf{k}}\rangle$ denote the eigenenergy and periodic part of the Bloch wavefunction, respectively, while $\hat{v}_{x}$ and $\hat{v}_{y}$ are the velocity operators along the corresponding Cartesian directions. The calculated Berry curvature along the high-symmetry path and its 2D distribution over the Brillouin zone are presented in Figs.~\ref{fig3}(d) and \ref{fig3}(e), respectively. Pronounced Berry-curvature contributions are observed in the vicinity of both valley regions and the zone center. In particular, the Berry curvatures at the $K$ and $K'$ valleys are approximately $2.9~\mathrm{\AA^2}$ and  $1.4~\mathrm{\AA^2}$, respectively, whereas a prominent contribution of opposite sign emerges at the $\Gamma$ point, reaching approximately $-3.1~\mathrm{\AA^2}$. Notably, the two valleys exhibit Berry curvatures with the same sign but distinctly different magnitudes. This distribution contrasts with the conventional valley-contrasting Berry-curvature behavior, in which the Berry curvature changes sign between $K$ and $K'$. The calculated values therefore establish an inequivalent, same-sign Berry-curvature distribution at the two valleys, indicating a valley-dependent momentum-space geometrical response.

The influence of Berry curvature on carrier dynamics can be understood through the semiclassical equations of motion. For an electron with charge $-e$, the anomalous contribution to its velocity under an in-plane electric field is given by~\cite{Xiao2010Berry}

\begin{equation}
\mathbf{v}^{\mathrm{anom}}_n(\mathbf{k})
=
\frac{e}{\hbar}\,
\mathbf{E}\times\boldsymbol{\Omega}_n(\mathbf{k}).
\end{equation}

Since the Berry curvatures at both valleys have the same sign, carriers originating from the $K$ and $K'$ valleys acquire anomalous transverse velocities with the same orientation but different magnitudes under an identical electric field. Consequently, the calculated Berry-curvature distribution does not support the conventional valley Hall mechanism based on oppositely directed transverse motion of carriers from the two valleys. Instead, it indicates a valley-dependent anomalous transverse response, with the magnitude of the carrier deflection depending on the valley from which the carriers originate, as schematically illustrated in Fig.~\ref{fig3}(f).

Beyond the Berry-curvature response, the substantial conduction-band valley splitting offers a possibility for gate-controlled valley selectivity. By tuning the chemical potential between the two conduction-band minima, electrostatic gating could preferentially populate the lower-energy valley under electron doping~\cite{Hossain2021GateValley,Zhang2024GateValley}. The finite Berry curvature of these states may additionally induce an anomalous transverse response, although the net Hall contribution depends on the Berry curvature of all occupied states. Thus, monolayer VBiSe$_3$ offers a potential platform for combining valley-selective carrier population with valley-dependent anomalous transverse dynamics, beyond conventional valley Hall transport.

\subsection*{Strain-induced quasi-half-valley-metallic state}

Strain engineering provides an effective route for tuning the electronic and valley properties of 2D materials~\cite{Sheoran2023Valley,Zang2021,Li2023}. We therefore investigated the response of monolayer VBiSe$_3$ to biaxial strain.

\begin{figure}[htp]
	\centering
	\includegraphics[width=0.5\textwidth]{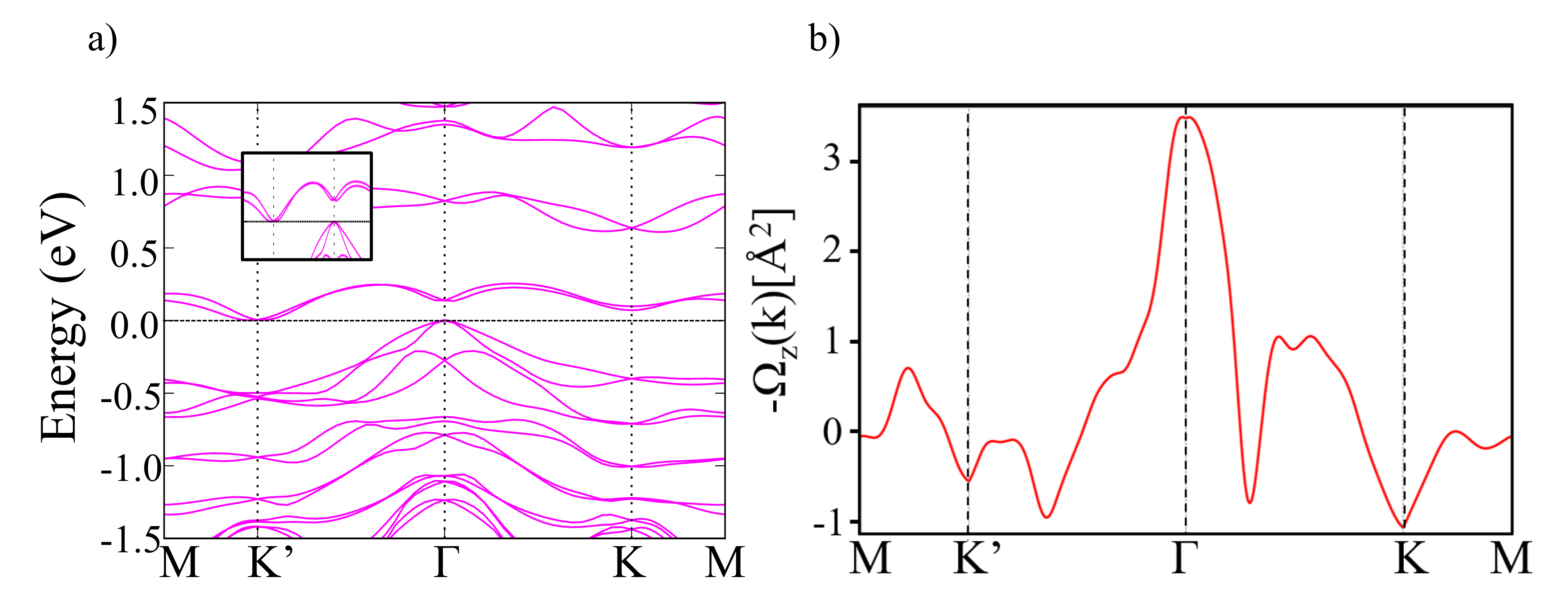}
	\caption{
		(a) Electronic band structure of monolayer VBiSe$_3$ under
		$-1\%$ biaxial compressive strain. The inset highlights the
		band-edge region, where the conduction-band minimum at $K'$
		and the valence-band maximum at $\Gamma$ approach the Fermi level.
		(b) Corresponding Berry curvature $\Omega_z(\mathbf{k})$
		along the high-symmetry path.
	}
	\label{fig4}
\end{figure}

A modest biaxial compressive strain of $\varepsilon=-1\%$ substantially modifies the band-edge configuration of monolayer VBiSe$_3$. As shown in Fig.~\ref{fig4}(a), the conduction-band minimum at the $K'$ valley and the valence-band maximum at the $\Gamma$ point simultaneously approach the Fermi level. This configuration is characteristic of a QHVM-like electronic structure, in which the electron-side band edge is valley-selective, whereas the hole-side band edge is not associated with the $K/K'$ valley degree of freedom~\cite{Guo2023QHVM}. Specifically, the electron-side band edge in VBiSe$_3$ is associated with the $K'$ valley, while the hole-side band edge is centred at $\Gamma$. A related strain-induced QHVM state has also been predicted in monolayer Cr$_2$CSF, although under unidirectional compressive strain~\cite{Phutela2025Cr2CSF}. These results indicate that biaxial compression provides an effective means of tuning the band-edge hierarchy in VBiSe$_3$ towards a QHVM-like regime.

The Berry-curvature distribution is also modified under compression, as shown in Fig.~\ref{fig4}(b). A pronounced Berry-curvature feature persists around the $\Gamma$ point, while the $K$ and $K'$ valleys continue to exhibit inequivalent Berry-curvature responses. The $\Gamma$-centred contribution is particularly relevant because it accompanies the hole-side band edge, whereas the electron-side band edge remains associated with the $K'$ valley. Thus, the strained system exhibits a momentum-dependent Berry-curvature distribution involving both the valley regions and the $\Gamma$-centred states. This observation highlights an additional aspect of the QHVM-like band-edge configuration in VBiSe$_3$, where the hole-side states also possess a non-negligible Berry-curvature contribution. Accordingly, the strained electronic structure provides a potential platform for studying valley-selective electron states alongside non-valley-centred hole states, although their resulting transverse transport properties would require further investigation.

Overall, a small biaxial compressive strain of $-1\%$ substantially reorganizes the band-edge hierarchy and modifies the Berry-curvature distribution of monolayer VBiSe$_3$, suggesting a route towards strain-controlled valley-selective electronic properties.

\subsection*{Conclusion}

In summary, using first-principles calculations and Wannier-based analysis, we have investigated the interplay of altermagnetism, SOC, and valley physics in monolayer VBiSe$_3$. In the absence of SOC, the N\'eel AFM state exhibits pronounced nonrelativistic momentum-dependent spin splitting with a characteristic $i$-wave texture. The inclusion of SOC reconstructs the electronic states, producing anisotropic Rashba-like spin splitting together with a sizable $K/K'$ valley splitting of approximately 112.2~meV. Orbital analysis and the effective SOC model indicate that the valley splitting originates from SOC acting on hybridized V-$d_{z^2}$/Se-$p_{x,y}$ states. Furthermore, the Berry-curvature distribution exhibits same-sign but unequal values at the $K$ and $K'$ valleys, implying valley-dependent anomalous transverse velocities with the same orientation but different magnitudes. This behavior distinguishes the system from conventional valley-contrasting Berry-curvature systems. We further demonstrate that a modest biaxial compressive strain of $-1\%$ substantially modifies the electronic structure, producing a QHVM-like band-edge configuration characterized by valley-selective electron states at $K'$ and non-valley-centred hole states at $\Gamma$. The associated Berry-curvature distribution also exhibits contributions from both the valley regions and the $\Gamma$ point. These findings establish monolayer VBiSe$_3$ as an interesting platform for exploring the interplay of compensated antiferromagnetism, SOC-driven spin splitting, and valley-dependent electronic properties, with strain offering an additional degree of freedom for controlling its electronic structure.

\section{Acknowledgement}
D. T. acknowledges UGC, India, for the senior research fellowship [Grant no. BUPME01420733/(CSIR-UGC-NET JUNE 2022)]. S. B. acknowledges financial support from SERB under a core research grant (Grant no. CRG/2019/000647) to set up his High Performance Computation (HPC) facility “Veena” at IIT Delhi for computational resources. We also acknowledge “Tejas” IoE facility of Department of Physics, IIT Delhi for HPC.

\bibliography{reference}
\end{document}


\renewcommand{\thefigure}{S\arabic{figure}}
\renewcommand{\thetable}{S\arabic{table}}
\renewcommand{\theequation}{S\arabic{equation}}
\begin{center}
    \Large \textbf{Coupled Altermagnetic and Valley Physics in Monolayer VBiSe$_3$} \\[1em]

    \normalsize
    Divyanshi Tyagi$^{*}$, Saswata Bhattacharya$^{\dagger}$ \\[0.5em]
    \textit{Department of Physics, Indian Institute of Technology Delhi, New Delhi 110016, India} \\[2em]

    \LARGE \textbf{Supplemental Material}
\end{center}

\vspace{2em}

\noindent \textbf{I.} Magnetic configurations and relative energies of monolayer VBiSe$_3$. \\

\noindent \textbf{II.} Dynamical stability of monolayer VBiSe$_3$: Phonon dispersion. \\

\noindent \textbf{III.} Orbital-projected electronic structure of monolayer VBiSe$_3$. \\

\noindent \textbf{IV.} Microscopic origin of valley splitting: Detailed spin--orbit-coupling Hamiltonian analysis. \\

\noindent $^{*}$ divyanshi.tyagi@physics.iitd.ac.in \\
\noindent $^{\dagger}$  saswata@physics.iitd.ac.in

\newpage

\section*{I. Magnetic configurations and relative energies of monolayer VBiSe$_3$}

\begin{figure}[h!]
\centering
\includegraphics[width=0.9\textwidth]{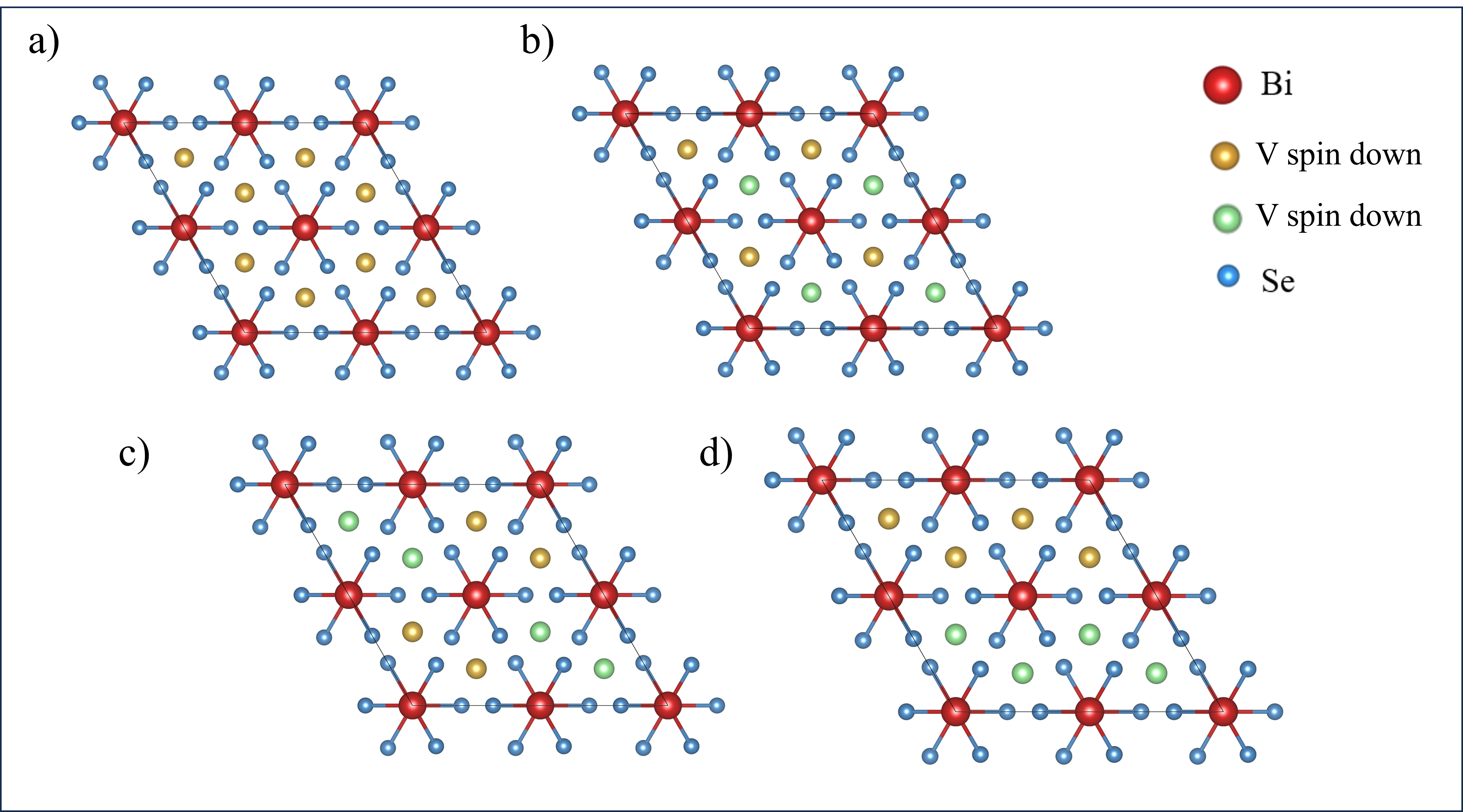}
\caption{Different magnetic configurations considered for monolayer
VBiSe$_3$. The red and blue spheres represent Bi and Se atoms,
respectively, while the golden and green spheres denote V atoms with
opposite spin orientations. The AFM-N\'eel configuration is found to be
the magnetic ground state.}
\end{figure}

At the PBE+$U$ level, the AFM-N\'eel state is taken as the reference
ground state. The AFM-stripy, AFM-zigzag, and FM configurations are
higher in energy by 134.18, 646, and 208~meV per unit cell,
respectively.
\newpage

\section*{II. Dynamical stability of monolayer VBiSe$_3$}

\begin{figure}[h!]
\centering
\includegraphics[width=0.6\textwidth]{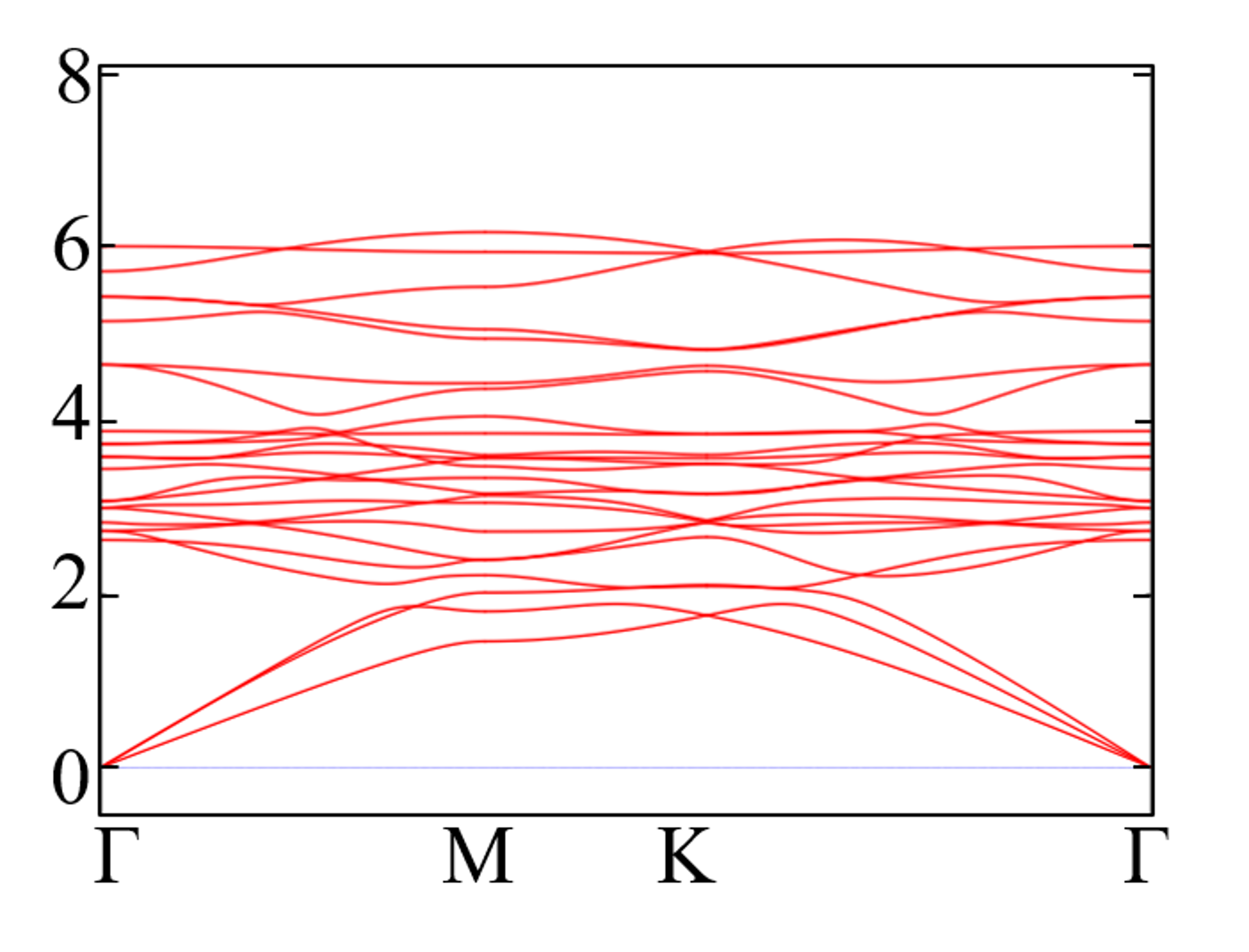}
\caption{Phonon dispersion of monolayer VBiSe$_3$ along the
$\Gamma$--M--K--$\Gamma$ high-symmetry path. The absence of significant
imaginary phonon frequencies confirms the dynamical stability of the monolayer.}
\end{figure}

\newpage

\section*{III. Orbital-projected electronic structure of monolayer VBiSe$_3$}

\begin{figure}[h!]
\centering
\includegraphics[width=1.05\textwidth]{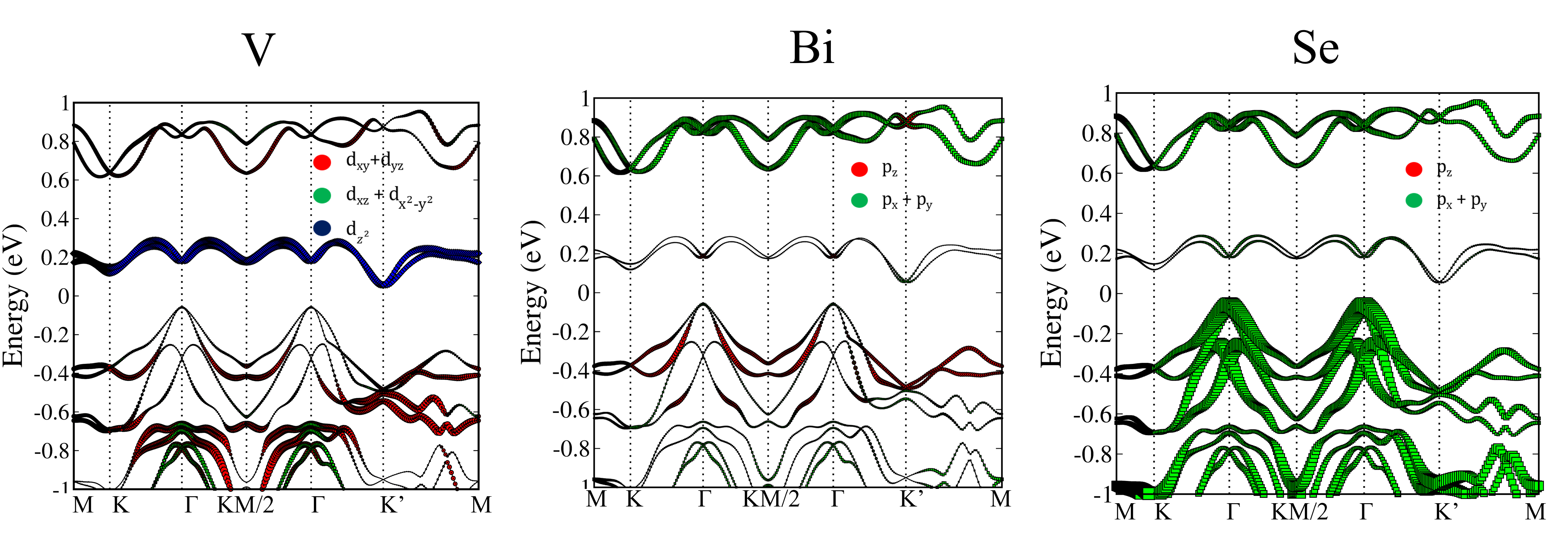}
\caption{Orbital-projected electronic band structures of monolayer
VBiSe$_3$ showing the contributions from (a) V-$d$, (b) Bi-$p$, and
(c) Se-$p$ orbitals. The valence-band edge is mainly dominated by
Se-$p_{x,y}$ states, whereas the conduction-band edge is primarily
derived from V-$d_{z^2}$ orbitals with finite Se-$p_{x,y}$ hybridization.}
\end{figure}

\section*{IV. Microscopic origin of valley splitting:
	detailed spin--orbit-coupling analysis}

To provide a more detailed microscopic description of the conduction-band
valley splitting in monolayer VBiSe$_3$, we consider the atomic
spin--orbit-coupling (SOC) Hamiltonian ~\cite{Dai2008SOC,Whangbo2015SOC}

\begin{equation}
\hat{H}_{\mathrm{SOC}}
=
\lambda\hat{\mathbf{S}}\cdot\hat{\mathbf{L}}
=
\hat{H}_{\mathrm{SOC}}^{0}
+
\hat{H}_{\mathrm{SOC}}^{1},
\end{equation}

where $\lambda$ is the SOC strength and
$\hat{\mathbf{S}}$ and $\hat{\mathbf{L}}$ are the spin and orbital
angular-momentum operators, respectively. Here,
$\hat{H}_{\mathrm{SOC}}^{0}$ and $\hat{H}_{\mathrm{SOC}}^{1}$ denote
the spin-conserving and spin-flip contributions, respectively. Since the
present analysis focuses on the leading valley-dependent energy shift
within a fixed spin channel, the spin-flip contribution
$\hat{H}_{\mathrm{SOC}}^{1}$ is neglected in the following first-order
analysis.

For an arbitrary orientation of the magnetic moment, specified by the
polar and azimuthal angles $(\theta,\phi)$, the spin-conserving part may
be written as

\begin{equation}
\hat{H}_{\mathrm{SOC}}^{0}
=
\lambda\hat{S}_{z'}
\left[
\hat{L}_{z}\cos\theta
+
\frac{1}{2}
\left(
\hat{L}_{+}e^{-i\phi}
+
\hat{L}_{-}e^{i\phi}
\right)
\sin\theta
\right],
\end{equation}

where $z'$ denotes the spin-quantization axis. For the out-of-plane
magnetic orientation obtained from the MAE calculation, $\theta=0$ and
$z'=z$, and the above expression reduces to

\begin{equation}
\hat{H}_{\mathrm{SOC}}^{0}
=
\lambda\hat{S}_{z}\hat{L}_{z}.
\end{equation}

The orbital-projected band structure shows that the conduction-band minimum
is predominantly derived from V-$d_{z^2}$ states with a finite contribution
from Se-$p_x$ and Se-$p_y$ orbitals. To isolate the SOC-active orbital
contribution, the Se-$p_{x,y}$ component of the conduction-band valley
states may be represented in the idealized orbital basis as

\begin{equation}
|\phi_c^\tau\rangle
=
\frac{1}{\sqrt{2}}
\left(
|p_x\rangle+i\tau|p_y\rangle
\right),
\end{equation}

where $\tau=+1$ and $-1$ denote the $K$ and $K'$ valleys, respectively.
Defining

\begin{equation}
\beta=
\langle\uparrow|\hat{S}_z|\uparrow\rangle ,
\end{equation}

the SOC correction to the $p$-orbital component of the conduction-band
valley state becomes

\begin{align}
E_c^\tau
&=
\langle\phi_c^\tau|
\hat{H}_{\mathrm{SOC}}^{0}
|\phi_c^\tau\rangle
\nonumber\\
&=
\frac{i\tau\lambda\beta}{2}
\left[
\langle p_x|\hat{L}_z|p_y\rangle
-
\langle p_y|\hat{L}_z|p_x\rangle
\right].
\end{align}

Using

\begin{equation}
\hat{L}_z|p_x\rangle=i|p_y\rangle,
\qquad
\hat{L}_z|p_y\rangle=-i|p_x\rangle,
\end{equation}

where $\hbar=1$, one obtains

\begin{equation}
E_c^\tau=\tau\lambda\beta.
\end{equation}

The SOC-induced energy difference between the two valley components is
therefore

\begin{equation}
\Delta E_c^{(p)}
=
E_c^{K}-E_c^{K'}
=
2\lambda\beta.
\end{equation}

Thus, the $p_x+ip_y$ and $p_x-ip_y$ components at the opposite valleys
carry opposite orbital angular momenta and experience opposite SOC energy
corrections.

It should be emphasized that the actual CBM of VBiSe$_3$ is not a pure
$p$-orbital state. As shown in the orbital-projected electronic structure
in Fig.~S3, the CBM is mainly composed of V-$d_{z^2}$ states. Since

\begin{equation}
\langle d_{z^2}|
\hat{L}_z
|d_{z^2}\rangle
=
0,
\end{equation}

a pure $d_{z^2}$ state cannot by itself generate a first-order
valley-dependent SOC shift. The finite Se-$p_{x,y}$ contribution therefore
provides the required SOC-active orbital character. Consequently, the
observed conduction-band valley splitting originates from SOC acting on
the hybridized V-$d_{z^2}$/Se-$p_{x,y}$ band-edge states rather than from
the V-$d_{z^2}$ component alone.

\bibliographystyle{apsrev4-2}
\bibliography{reference}